\documentstyle[12pt,epsf,epsfig]{aipproc}

\begin{document}
\sloppy         
\title{\LARGE\bf An effective method to read out large \\
 scintillator areas with precise timing}
\author{J. B\"ahr, H.-J. Grabosch, V. Kantserov\footnote{On leave from
    Moscow Physics Engineering Institute}, H.Leich,\\ R.Leiste,
  R.Nahnhauer}
\address{DESY-Zeuthen, 15738 Zeuthen, Germany}
\maketitle
\begin{abstract}
        Using scintillator tile technology several square meters 
        of plastic scintillator are read out by only two
        photomultipliers with a time precision of about 1.5 nsec.
          Two examples are discussed to build a detector based on this
        technology to search for cosmic muons and neutrinos.
\end{abstract}

\section*{Introduction}
The readout of plastic scintillators is done classically using
adiabatic clear light guides to focus the produced photons to the
photocathode of a photomultiplier (PM). Later wavelength shifting rods
were applied to collect the light of larger
scintillator areas. In recent years wavelength shifting fibers
have been used to read out scintillator tiles of preshowers or
calorimeter moduls [1,2,3].

Precise timing normally demands a large number of photons to be
produced in the scintillator. Using two
photomultipliers at both ends of not too large scintillator rods
and a meantimer for the output signals a time precision of a few
hundred psec is in reach. For large scintillator areas this method is
rather expensive because many photomultipliers are needed.
  Using wavelength shifting fibers for the scintillator read out has
the clear disadvantage that only a few percent of the produced light
will be trapped to the PM photocathode. Nevertheless we will propose
in the following a method to read out large detector areas with a
few photomultipliers retaining a time precision of about 1 nsec.  

\section*{The measuring principle}
\begin{figure}[t] 
\vspace*{3cm}
\begin{center}
\begin{minipage}[t]{10cm}
\epsfig{file=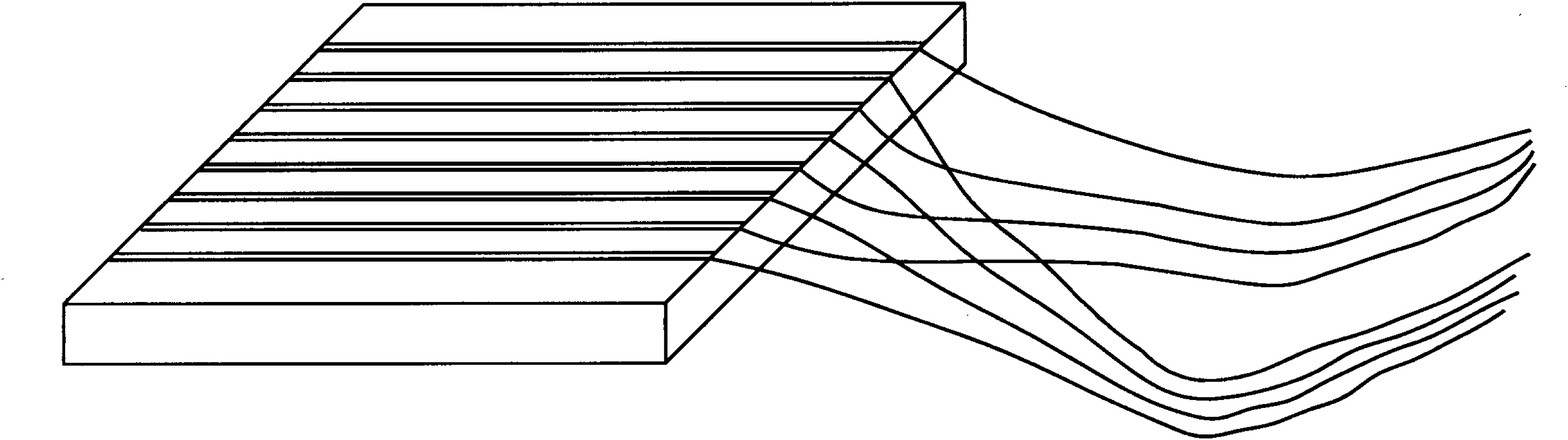,width=10cm}
\caption{Schematic view of a scintillator tile readout by
 wavelength shifting fibers.}
\label{bild1}
\end{minipage}
\end{center}
\hfill
\vspace*{3cm}
\begin{center}
\begin{minipage}[b]{10cm}
\epsfig{file=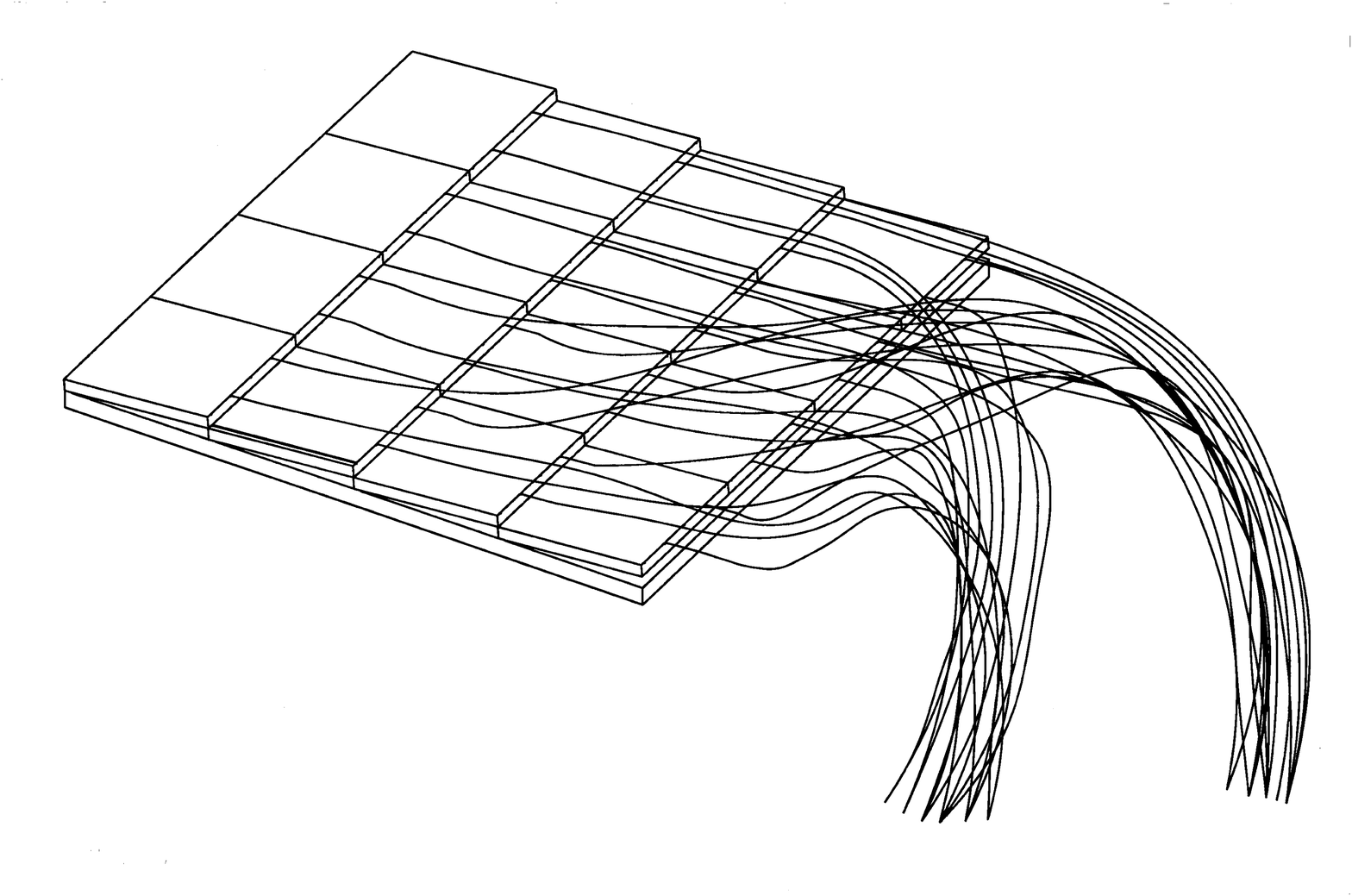,width=10cm}
\caption{Combination of scintillator tiles to cover larger
 areas. Fibers from all tiles are splitted in two groups
for the read out.}
\label{bild2}
\end{minipage}
\end{center}
\end{figure}

The basic piece of a large scale detector is a small scintillator
tile of about ~25 x 25 x 2 cm$^3$ as schematically drawn in
fig. \ref{bild1}. Two
groups of wavelength shifting fibers are glued into grooves at it's
surface. These fibers are connected to clear optical fibers of a
length of several meters guiding the light produced by crossing
particles to two photomultipliers. A particle hit is registrated if
both PM's give a coincident signal above a certain threshold and within a short time window. This
demand reduces already the PM noise by orders of magnitude.

Several scintillator tiles are combined by summing up the clear 
read out fibers of each of the two groups (see fig. \ref{bild2}). A natural
restriction is given only by the size of the photocatode of the
used PM's in relation to the number and size of the fibers per tile.
For a 2'' PM with homogeneous response of the whole photocathode
about 100 tiles can be combined to more than 6 m$^2$ detector area
if four fibers of 1.5 mm diameter per tile and group are used.

\section*{Results for single tiles}
For the construction of tiles we used different types of
scintillator of 1 - 2 cm thickness.The fibers tested came mainly from
BICRON\footnote{BICRON, 12345 Kinsman Road, Newbury, Ohio, USA}. 
Finally we used the fast wavelength shifting fiber BCF 92
with a decay constant of 2.7 nsec together with 2 m long clear fibers
BCF 98 both with 1.5 mm diameter.

Several fiber arrangements were tested using between 2 and 8 fibers
per group. An optimum was found for four fibers per group equally
distributed over the tile surface. The fibers were coupled to two
Philips\footnote{Philips Photonique, Av. Roger
  Roacier, B. P. 520, 19106 Brive, France} photomultipliers XP 2020 which were read out by charge sensitive
ADC's. Two additional scintillation counters on top
and bottom of the tile studied are used to trigger for throughgoing
cosmic particles. The efficiencies derived from measured ADC-spectra 
are given in the table for a 1 cm thick tile made from BICRON 408 
together with those for a 2 cm
thick scintillator produced at IHEP Serpukhov. In both cases values
near to 99 \% are found for the tile hit efficiency. The light output
increases by about a factor 1.5 if double clad fibers are used.  

\begin{table}[h!]
\caption{Number of photoelectrons N$_{pe}$ and efficiencies
  $\varepsilon$ for two scintillator tiles read out by two groups of
  single clad fibers BCF~92 coupled to 2 m long clear optical fibers
  BCF~98}
\label{tabelle}
\renewcommand{\arraystretch}{1.5}
\begin{tabular}{|c|c|c|c|c|c|}  \hline
scintill. & d, cm & N$_{pe}^{(1)}$ & N$_{pe}^{(2)}$ & $\varepsilon_1$
&$\varepsilon_2$ \\ 
\hline
BC 408    &   1    &   4.7  &   4.7  &  0.991&  0.991 \\ \hline
Serpukhov &    2   &     4.6&     4.4 &     0.990 & 0.998 \\
 \hline
\end{tabular}
\end{table}

\noindent
Keeping in mind, that the two time measurements are independent of each
other we derive from their difference a time resolution of $\sim$ 1.5
nsec for the single channel signal.
The timing behaviour of the scintillator signals was studied using
TDC's which provide a time resolution of 100
psec. The TDC's are read out in common stop mode. The stop signal is
derived from a coincidence of the signals of the two fiber groups of a
tile using very low thresholds of about 5 mV.

\section*{First applications}
\subsection*{A t$_0$--detector for the L3COSMIC-Project}
Two years ago the idea was born to use this technology for a  $~$ 50 m$^2$
scintillation detector to be installed on top of the L3 experiment at
CERN\footnote{see http://hpl3sn02.cern.ch/l3\_cosmics/}. This will allow to use the muon drift chambers of this
experiment to measure cosmic particle momenta because a precise
$t_0$-signal will be available. As discussed already since long
\cite{lit4} a rich spectrum of cosmic particle physics could be investigated
in this L3COSMIC experiment. The time resolution demanded for a $\pm$
1 \% measurement of the cosmic muon momentum spectrum between 20 GeV and 1 TeV is
about 1 -- 2 nsec just in reach for the proposed device.

The detector will consist of 8 modules of 6 m$^2$ each. The first
module has been constructed at DESY-Zeuthen and tested in detail
with cosmic particles. The tiles have a size of
25 x 25 x 2 cm$^3$, therefore 96 of them are needed for a full size 
module. Half of the tiles were produced from Serpukhov 
scintillator, the other half from already used
one delivered by the University of Michigan\footnote{We thank L.B. Jones
  from the University of Michigan for providing the
corresponding material.}. For the read out 1.5 mm double
clad fibers BCF 92 and BCF 98 were taken.
 
\begin{figure}[t] 
\begin{minipage}[t]{7cm}
\epsfig{file=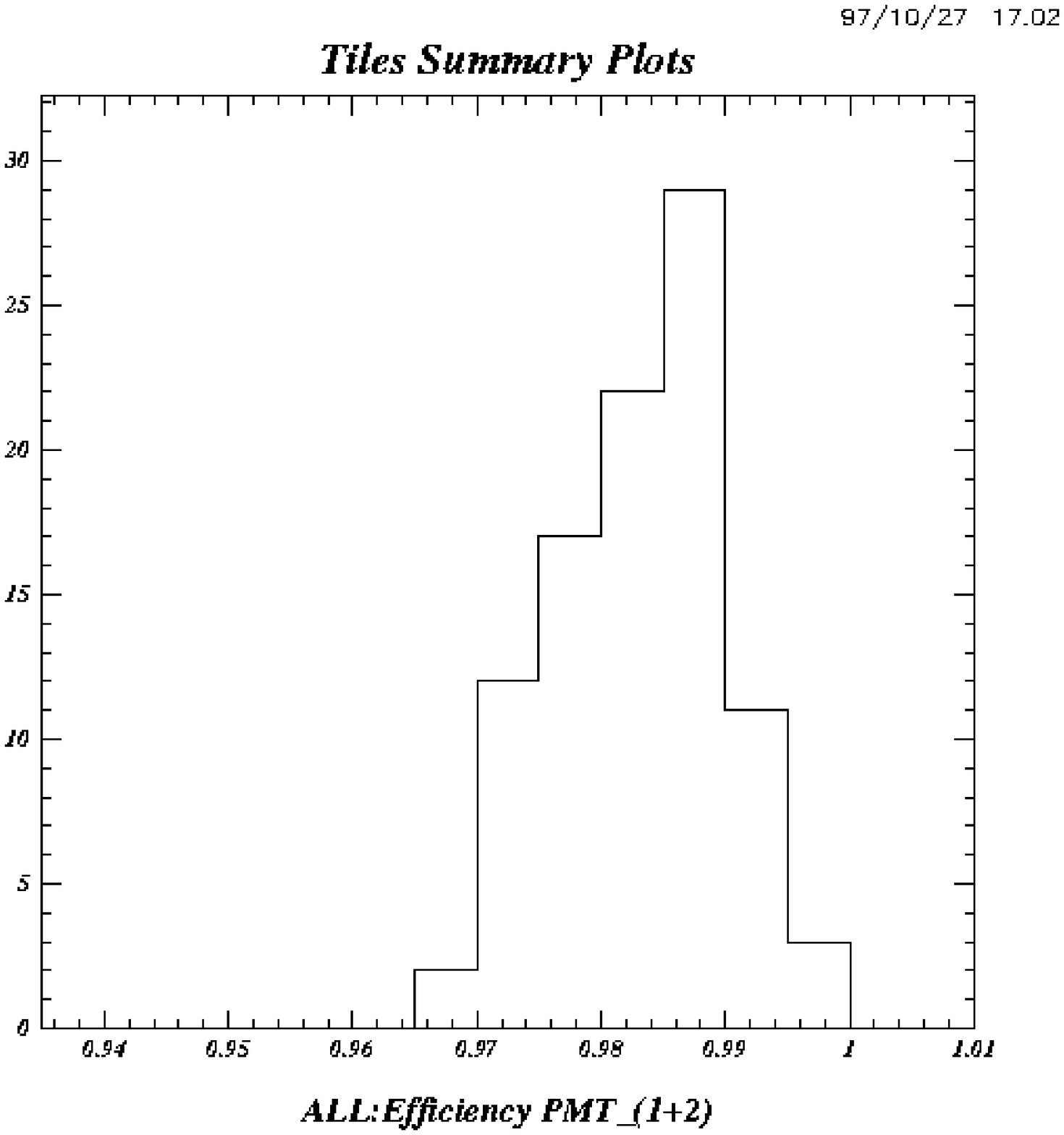,width=7cm}
\caption{Efficiency distribution for 96 tiles of the first module
 of th L3COSMIC t0-detector.}
\label{bild3}
\end{minipage}
\hfill
\begin{minipage}[t]{7cm}
\epsfig{file=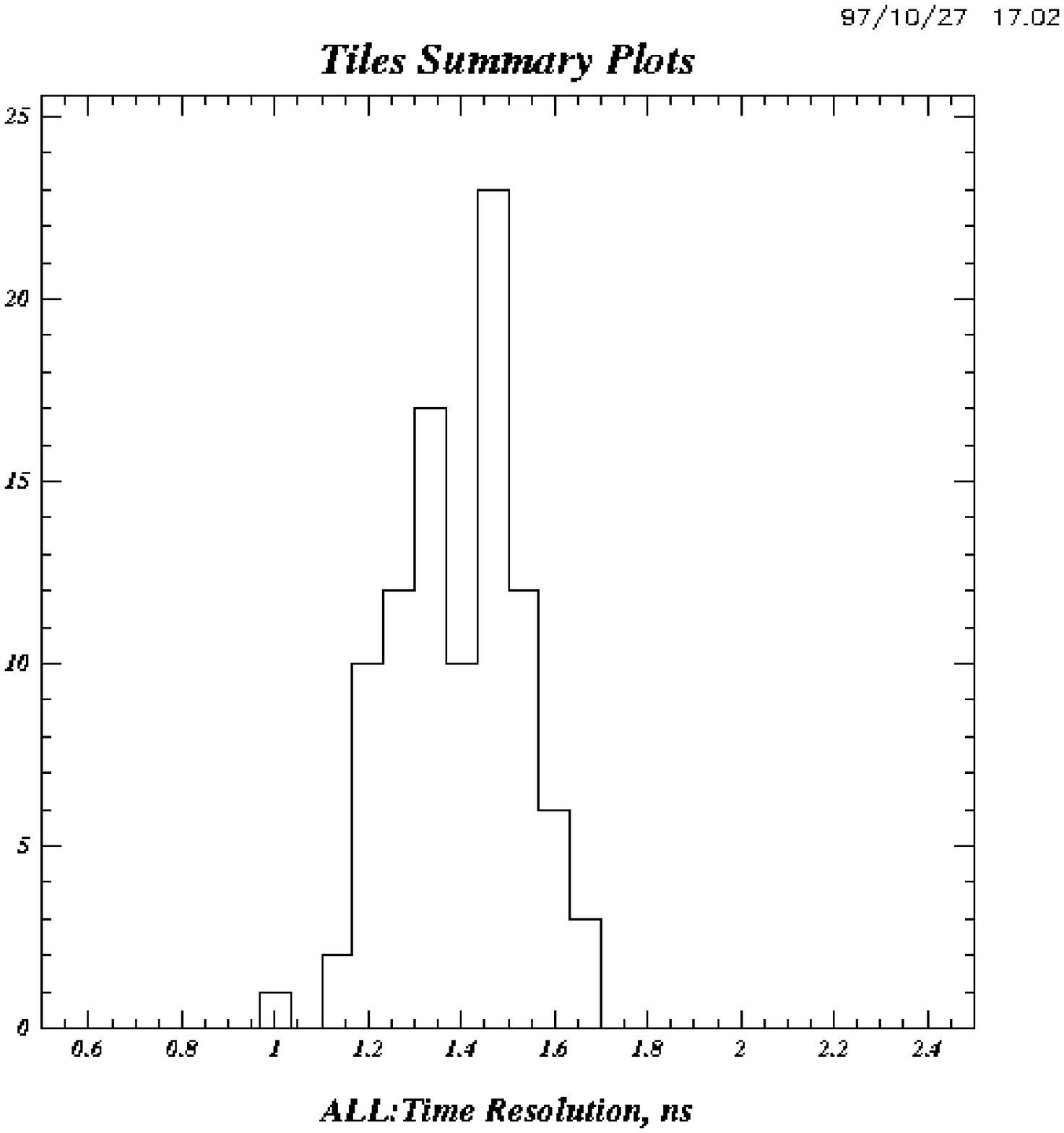,width=7cm}
\caption{Time resolution distribution for 96 tiles of the first
 module of the L3COSMIC t0-detector.}
\label{bild4}
\end{minipage}

\hfill
\begin{minipage}[b]{7cm}

\vspace*{-1cm}
\epsfig{file=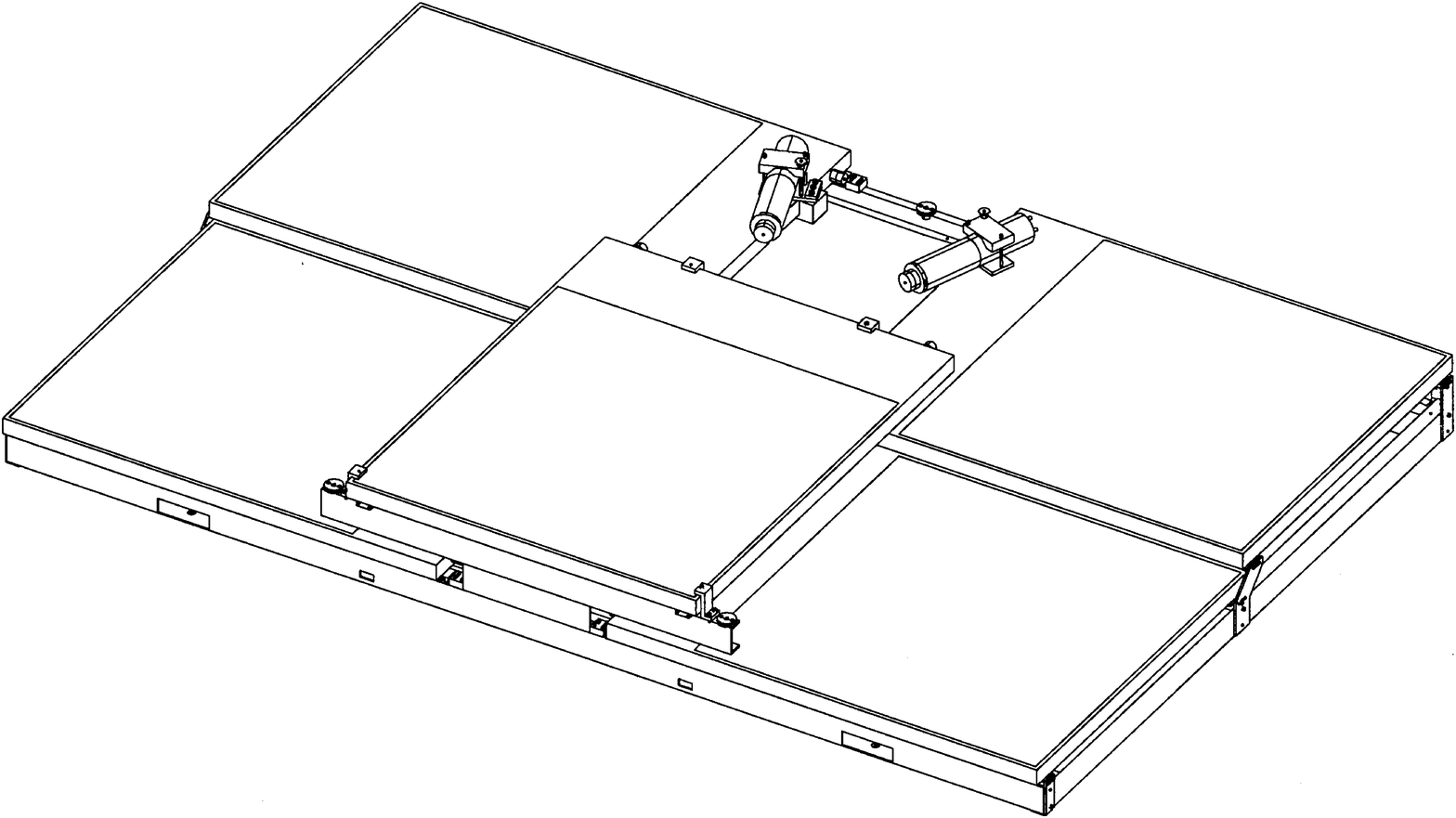,width=6cm}
\vspace*{1cm}
\caption{Mechanical layout of a full size 6 m$^2$ module readout
 with two photomultipliers.}
\label{bild5}
\end{minipage}
\hfill
\begin{minipage}[b]{7cm}
\epsfig{file=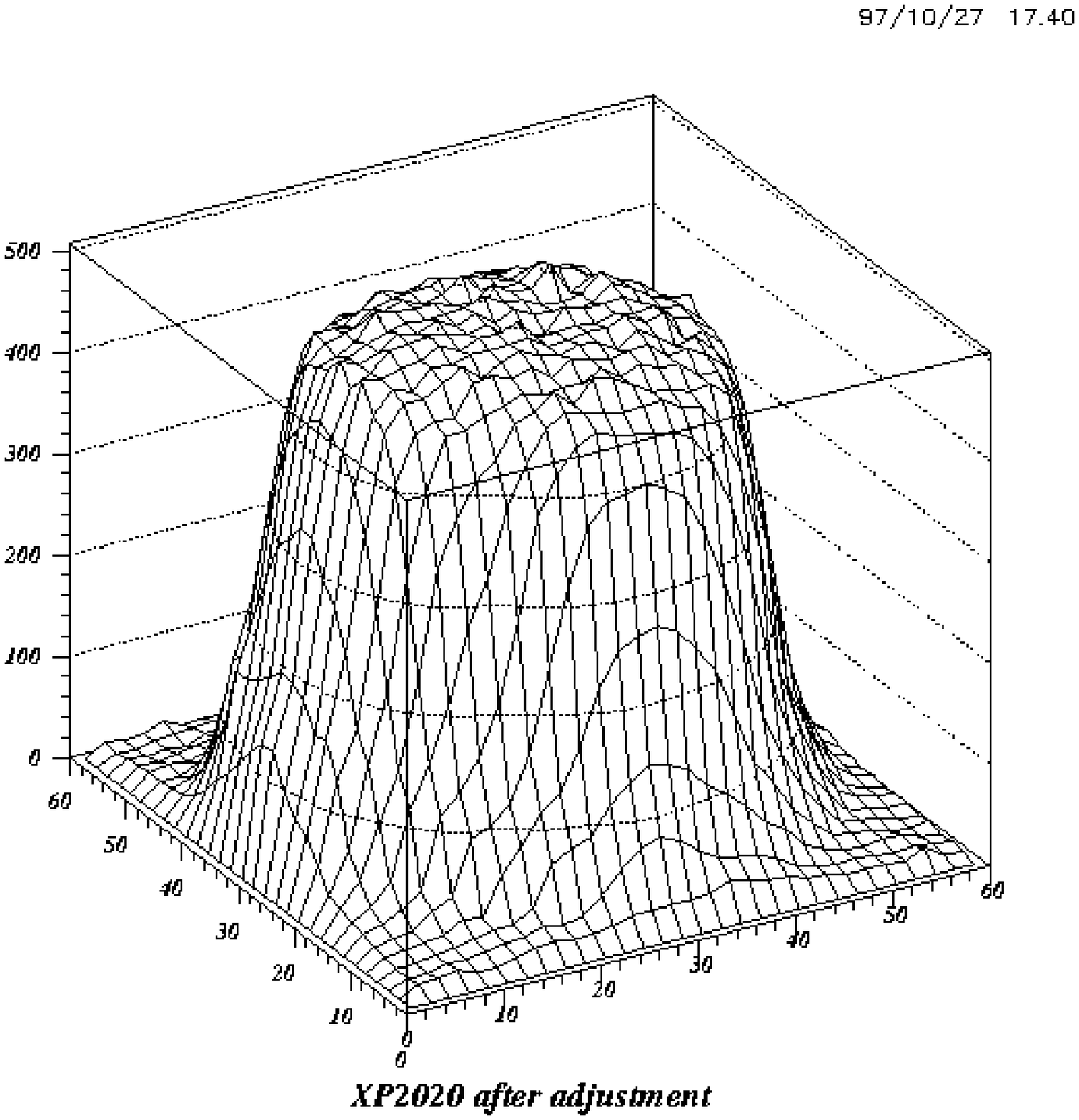,width=6cm}
\caption{Example for the sensitivity distribution of an XP2020
 photomultiplier across its entrance window surface.}
\label{bild6}
\end{minipage}

\end{figure}

The results of our measurements are given in figs. \ref{bild3} and
\ref{bild4} for the
hit efficiency and the time resolution respectively. Average values of

\begin{center}
$\varepsilon_{12}$ = 98.3 $\pm$ 0.7 \%
and  
$\sigma_t$  = 1.4 $\pm$ 0.1 nsec
\end{center}

\noindent
demonstrate that we have reached the design goal.

Each 16 tiles are dense packed inside an aluminium box to cover
an area of 1 m$^2$. Six of the boxes are assembled together to form a full
size module (see fig. \ref{bild5}). The geometrical arrangement is done in a way
that a 6 m$^2$ area is filled without gaps allowing to put the two times
6 fiber bundles to the two XP 2020 photomultipliers for the read out. 
Practically this coupling is done using a special mask arrangement
Because the mask covers the whole surface of the PM's
photocathode one has to make sure equal sensitivity of the PM
independent of the fiber position. The high voltage divider used
allows a correponding adjustment. The result is presented for one PM
in fig. \ref{bild6}.
  The full size module has been tested first in Zeuthen using small
(10 x 10 cm$^2$) scintillator paddles to allow position dependent cosmic
particle triggering. We found an average hit efficiency of 99.1 $\pm$
0.4 \% and
an average time resolution of 1.4 $\pm$  0.1 nsec.      

Two testruns have been performed in autumn 96 and spring 97 at
CERN. A 3 m$^2$ and a 6 m$^2$ detector respectively were installed  on top
of the L3--magnet. Data were taken in coincidence with the barrel
scintillator system of L3. Two TDC's of this system and the L3-DAQ were
used to measure the arrival time of crossing particles  as seen by the
PM's of the t$_0$-detector module. With this arrangement we got a time
resolution of $\sigma_t$ = 1.5 $\pm$ 0.1 nsec which 
confirms the laboratory results. 

\subsection*{A Scintillator based cosmic Particle detection Yard (SPY)}
The read out scheme described in section 2 has first been proposed
for a cosmic neutrino detector project at earth surface. The aim was
twofold. First, we wanted to detect high energetic air showers from
above as well as "astrophysical" neutrinos from below the
ground. Second, easy construction and installation was demanded
allowing 100 \% access for maintainance and exchange of components 
avoiding the extreme boundary conditions of present cosmic neutrino
detection techniques [5,6].  

\begin{figure}[b] 
\begin{minipage}[t]{7cm}
\begin{center}
\epsfig{file=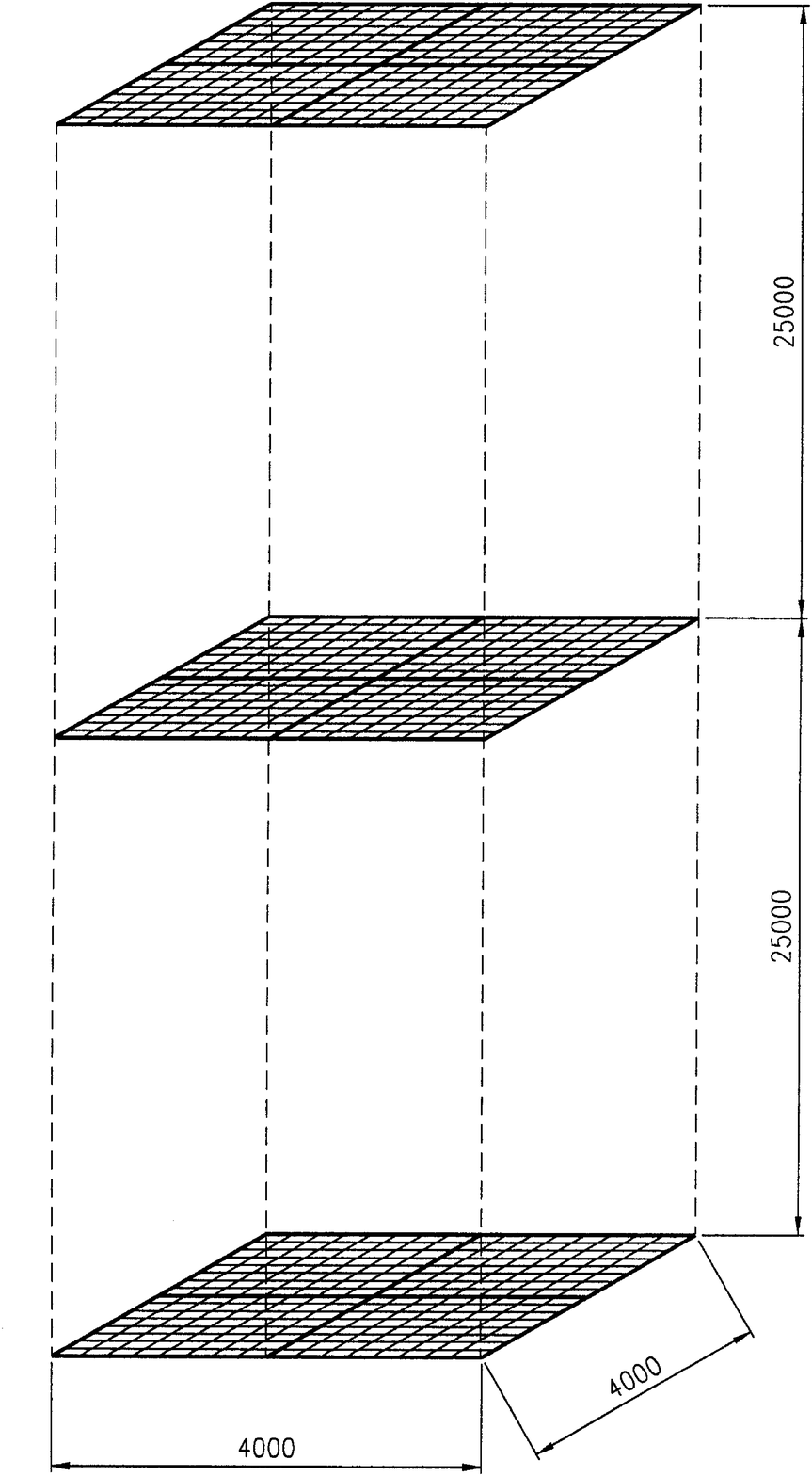,width=4cm,}
\end{center}
\caption{Schematic view of a three-floor SPY tower with 4x4 m$^2$
 scintillator sheets made out of scintillator tiles.}
\label{bild7}
\end{minipage}
\hfill
\begin{minipage}[t]{7cm}
\epsfig{file=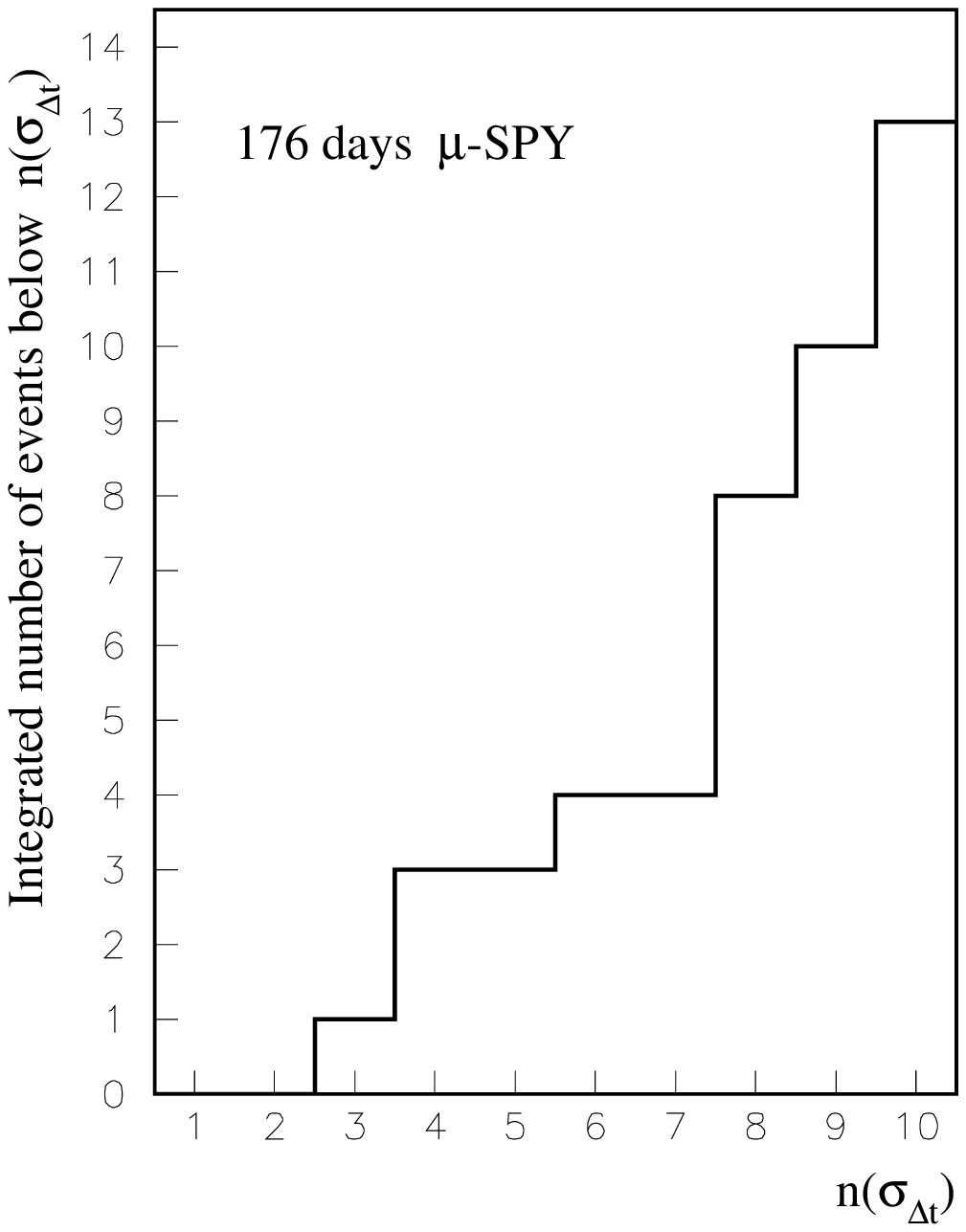,width=7cm}
\vspace*{0.0cm}
\caption{Distribution of tracks (in bins of the time resolution of
             3 nsec)  for the radial difference to a possible
             neutrino signal from below the ground in the
             $\Delta t_{21}-\Delta t_{32}$ plane for 176 days of data taking with $\mu$-SPY}
\label{bild9}
\end{minipage}
\hfill
\vspace*{-2cm}
\begin{center}
\begin{minipage}[b]{12cm}
\epsfig{file=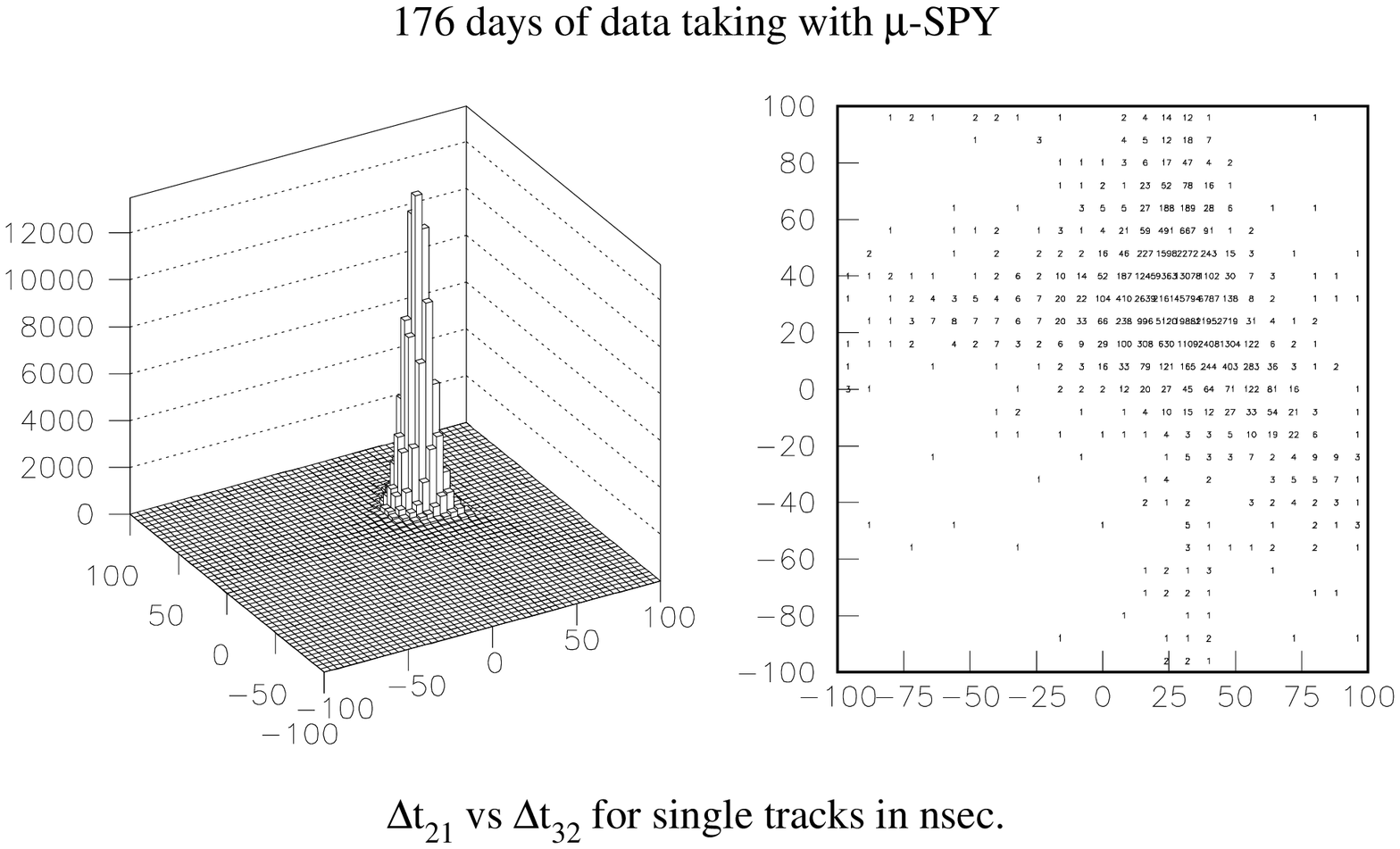,width=10cm}
\caption{Distribution of tracks in the plane of time differences
 $\Delta t_{21} vs.\Delta t_{32}$ for cosmic particles crossing $\mu$-SPY.}
\label{bild8}
\end{minipage}
\end{center}

\end{figure}

The idea was to use modern scintillator technology to build a fast
detector, well structured to keep the tremendous background from
normal cosmic rays small. Basic elements of such a detector could be
scintillator sheets made out of tiles forming a three floor tower (see
fig. \ref{bild7}). Defining a hit as a coincidence of the two photomultipliers
of one sheet as described above, a track is identified by a delayed
coincidence of hits of all three planes. A sheet size of 4 x 4 m$^2$
would have to cope with a cosmic ray rate of about 4 kHz. A distance of
25 m between every two sheets seems to allow to do that without
demanding an extreme time precision from the scintillator signal. 
The background rate per tower would be reduced to 6.4 Hz.

A large area detector could be build grouping many towers dense to
each other. With 25 x 25 towers of the described size a total area of
10 000 m$^2$ is reached. The detector could measure naturally also
inclined tracks. In this case the expected time delay would depend on
the particle direction. Predicting hit positions in the second plane
from those in the first and the last one would allow to keep the
background still limited to the extension of one sheet size. 

Monte Carlo calculations supported the above concept. However one
has to handle a background rate of about 10$^{11}$ per year which is
difficult to simulate.  Therefore we decided to build a small scale
prototype detector ($\mu$--SPY) at DESY-Zeuthen. Three
planes of 1 m$^2$ size divided in four subsections were installed in a
distance of about 10m between every two of them. The fibers of each
plane were read out with eight channels of a sixteen channel Philips
 R4760 photomultiplier. The time resolution per plane was measured
comparing hit arrival times of multihit triggers. It was found to be
about 3 -- 4 nsec and depends mainly on the quality of the
used PM. For the data acquisition we used a OS9 based
VME--system. The hit data were
collected in a tandem buffer. If full, it was transfered
to a SUN workstation where track reconstruction was done online. A
data reduction factor of 10$^4$ was reached.

In 176 days we observed 4.8 x 10$^9$ hits. After the online filter we
found 281.553 events with 324.034 tracks.The time differences between
planes one--two and two--three are shown in correlation to each other in
fig. \ref{bild8}. A clear enhancement is observed for normal cosmic rays
crossing the detector from above. A very small number of hits is found
in the opposite direction, as can be seen from fig. \ref{bild9}. Assuming
the worst case, a flat background in the region of hits from neutrino
interactions from below, we find a density of $\rho_{bg}$ =
0.009 events/1 nsec$^2$ for 1 m$^2$ and 1 year of running. Extrapolating
this number to a 10 000 m$^2$ detector we would get a signal to noise ratio of
S/N = 0.5 for atmospheric neutrinos above 2 GeV. We have to show
however, that the hit prediction for the second plane works for a
detector of this size.

\section*{Summary}
It has been demonstrated, that small size scintillator tiles read out with
two bundles of wavelength shifting fibers coupled by clear fibers of
several meters length to standard photomultipliers allow to detect
the crossing of minimum ionizing particles with about
99 \% efficiency and a timing precision of 1.5 nsec.

A dense pack of many of these tiles gives the possibility to read
out considerable large detector areas with only a few
photomultipliers. Because all fibers of all tiles have the same
length the properties of the detector are completely determined by the
single tile behaviour independent of the position of a crossing
particle.

\vspace{1cm}
{\bf Acknowledgement}\\
We want to thank our colleagues K.H.Sulanke and G.Trowitzsch
for their contribution to electronics and online software.

The testruns at CERN would not have been possible without the
support of J.J. Blaising, P. Le Coultre and U. Uwer. Their help is
gratefully acknowledged.

The SPY-project took profit from a lot of discussions with our
colleagues of the Baikal-Amanda group at DESY-Zeuthen.

\clearpage

\newpage

\clearpage

\newpage

\clearpage

\newpage

\begin{references}
\bibitem{lit1}Bamberger, A., et al., {\it NIM} {\bf A382}, 419, (1996)
\bibitem{lit2}Blair, R., et al., FERMILAB-PUB-96-390E, (1996)
\bibitem{lit3}For present projects see also:\\
       Freeman, J., contribution to this workshop\\
      Para, A., contribution to this workshop
\bibitem{lit4}B\"ahr, J., et al., L3 internal note 1977, (1996)
\bibitem{lit5}Belolaptikov, I.A., et al., {\it Astropart. Phys.} 
              {\bf 7}, 263, (1997)
\bibitem{lit6}Lowder, D.M., et al., Proceedings of the 17th Intern.Conf.
       on Neutrino Physics and Astrophysics, ed. K.Enquist,
       K.Huitu, J.Maalampi, Helsinki, Finland, 1996, P. 518
\end{references}
\end{document}